\documentclass[5p,times,procedia]{elsarticle}
\usepackage{ecrc}
\usepackage{setspace}

\volume{00}

\firstpage{1}

\journalname{Procedia CIRP}

\runauth{Zhibo Zhou et al.}

\jid{trpro}

\usepackage{amssymb}
\usepackage[figuresright]{rotating}
\usepackage{hyperref}
\hypersetup{colorlinks,allcolors=black}

\begin{document}
\begin{frontmatter}



\dochead{57th CIRP Conference on Manufacturing Systems 2024 (CMS 2024)}%

\title{Towards learning digital twin: case study on an anisotropic non-ideal rotor system}


\author[a]{Zhibo Zhou \corref{*}} 
\author[a]{Michael Walther}
\author[b]{Alexander Verl}
%\ead{author@institute.xxx}

\address[a]{Robert Bosch GmbH, Robert Bosch Campus 1, Renningen, 71272, Germany}
\address[b]{Institute for Control Engineering of Machine Tools and Manufacturing Units, University of Stuttgart, Seidenstraße 36, Stuttgart, 70174, Germany}

\aucores{* Zhibo Zhou. {\it E-mail address:} zhibo.zhou@de.bosch.com}

\begin{abstract}
In the manufacturing industry, the digital twin (DT) is becoming a central topic. It has the potential to enhance the efficiency of manufacturing machines and reduce the frequency of errors. In order to fulfill its purpose, a DT must be an exact enough replica of its corresponding physical object. Nevertheless, the physical object endures a lifelong process of degradation. As a result, the digital twin must be modified accordingly in order to satisfy the accuracy requirement. This article introduces the novel concept of "learning digital twin (LDT)," which concentrates on the temporal behavior of the physical object and highlights the digital twin's capacity for lifelong learning. The structure of a LDT is first described. Then, in-depth descriptions of various algorithms for implementing each component of a LDT are provided. The proposed LDT is validated on the simulated degradation process of an anisotropic non-ideal rotor system.   
\end{abstract}

\begin{keyword}
Digital twin \sep Lifelong learning \sep Model calibration




\end{keyword}
% \cortext[cor1]{Zhibo Zhou. Tel.: +49(711)811-30136}

\end{frontmatter}

% \linenumbers


\section{Introduction}
% \begin{figure}
% 	\centering
% 	\includegraphics[width=1.0\linewidth]{digital twin.PNG}
% 	\caption{The structure of a digital twin \cite{b7}} 
% 	\label{fig:digital twin}
% \end{figure}

A digital twin (DT) refers to a comprehensive physical and functional description of a component, product, or system \cite{b4}. DT has been a central topic in manufacturing because it can be used to improve machines' situational awareness, enhance operation resilience and flexibility, help design human-centered human machine collaboration strategies to increase the physical and psychological health of workers, allow the establishment of a self-organizing factory environment, provide unprecedented visibility into operation performance, and create the possibility of predicting future needs \cite{b8}. DT is regarded as a potential solution to increase automation and advance towards smart manufacturing \cite{b28}.

The principal component in a DT is a computational model of the physical twin \cite{b10}. However, the physical counterpart can suffer from a lifelong degradation. Therefore, it is necessary to make corresponding adjustments to the DT so that the degradation is also reflected on the DT and its influence is taken into account in the simulation for the behavior of the physical twin. This characteristic of DT is referred to as "self evolution" in \cite{b9, b26}. In \cite{b16}, the adjustment of a DT of robotic work-cells is realized through a multilevel calibration method. The generated DT can ensure sufficient accuracy for the offline planned robotic operations. In \cite{b18}, the adaptation of the model of rotating machinery is realized based on particle swarm optimization. The constructed DT-rotor model enables accurate simulation. In \cite{b10}, an evolving DT is proposed and verified on a system consisting of two cascading tanks driven by a pump. The property of "evolving" is realized based on the adaptation of weighting parameters in DT. \cite{b19} realized the modification of the model for dynamic flotation process simulation.

Since the physical counterpart changes constantly throughout life, the adaptation of DT should be activated multiple times. \cite{b7} state that suitable decision-making mechanisms are required, such that when a change on the physical counterpart is monitored, the corresponding modification process of the DT can be enacted. The window-based method is the most common method regarding adaptation in the lifelong of the machine \cite{b10,b19}. Window-based detection refers to the detection and adaptation of DT that are conducted over a period of time. The detection can only be triggered at the beginning of each period. Hence, the performance of this kind of adaptation method depends largely on the choice of the length of the window \cite{b11}.  Consequently, Window-based detector is a \textit{passive} detector. In comparison, an \textit{active} detector refers to some algorithm in which the adaptation is activated only when a change in the physical twin is first detected. 

Active methods have been successfully conducted to detect changes in the target concept in the data stream, which is known as concept drift \cite{b1}. Several different methods have been proposed \cite{b13, b14, b15}. \cite{b2} reviews the performance of different detectors for classification, while \cite{b3} reviews the performance of detectors for regression. However, there is currently little research regarding concept drift detection for DT, where the target concept is a dynamical system instead of a static distribution.

In this paper, inspired by successful detectors in concept drift, two kinds of detectors to detect changes in a dynamical system are proposed. Their performances are verified and compared on a constructed DT. The contribution of this paper is the following:
\begin{itemize}
    \item The concept of a learning digital twin (LDT) is proposed and implemented.
    \item The "learning ability" of different models are verified.
    \item Active detectors inspired by methods in "concept drift" are implemented to realize the modification of DT.
\end{itemize}

This paper is structured as following. A concrete description of method in Section \ref{sec3}. The result of a simulation experiment is given in Section \ref{sec4}. In the end, a conclusion is given in Section \ref{sec5}.

% \begin{equation}
% \begin{array}{lcl}
% \displaystyle X_r &=& \displaystyle\dot{Q}^{''}_{rad}\left/\left(\dot{Q}^{''}_{rad} + \dot{Q}^{''}_{conv}\right)\right.\\[6pt]
% \displaystyle \rho &=& \displaystyle\frac{\vec{E}}{J_c(T={\rm const.})\cdot\left(P\cdot\left(\displaystyle\frac{\vec{E}}{E_c}\right)^m+(1-P)\right)}
% \end{array}
% \end{equation}

%######################   Methodology     ###############################################################################################
\section{Methodology}
\label{sec3}
In this part, a detailed description of LDT is provided. Firstly, the problem referring to as concept drift for DT is proposed and defined in section \ref{sec3_1}. Then an overview of the structure of a LDT is shown in section \ref{sec3_2}. After that, a further detailed description of each block and the used method are given in section \ref{sec3_3} to \ref{sec3_5}.

\subsection{Concept drift for DT}
\label{sec3_1}
In \cite{b23} a concept drift is defined as following: between time point $t_0$ and time point $t_1$:
\begin{equation}
    \exists X: p_{t_0}(X,y)\neq p_{t_1}(X,y),
\end{equation}
where $p_{t_0}$ denotes the joint distribution at time $t_0$ between the input variable $X$ and the target variable $y$. In the context of DT, input variables $X=[u,x]$ consists of two parts, the system input $u \in \mathbb{R}^p $ and system state $x \in \mathbb{R}^n$, while target variable $y \in \mathbb{R}^q$ refers to the system output. 

In the theory of concept drift, it is of interest to distinguish between virtual drift ($p(X )$ changes) and real drift ($p(y|X)$changes). However, in LDT it is real drift that matters since it indicates the change of the physical object. In this paper, we focus on real drift where $p(y|X)$changes. 
%########################
\subsection{LDT}
\label{sec3_2}
Learning refers to the adaptation of the behavior of the DT so that it corresponds with the measured data and can accurately predict the future behavior of the physical object. It targets solving the decreased accuracy of DT when the physical object suffers from performance degradation. The concept of LDT is proposed so that the learning ability of the DT can be automatically activated with minimal adaptation times over the life of the physical object. 

\begin{figure}[h]
	\centering
	\includegraphics[width=1.0\linewidth]{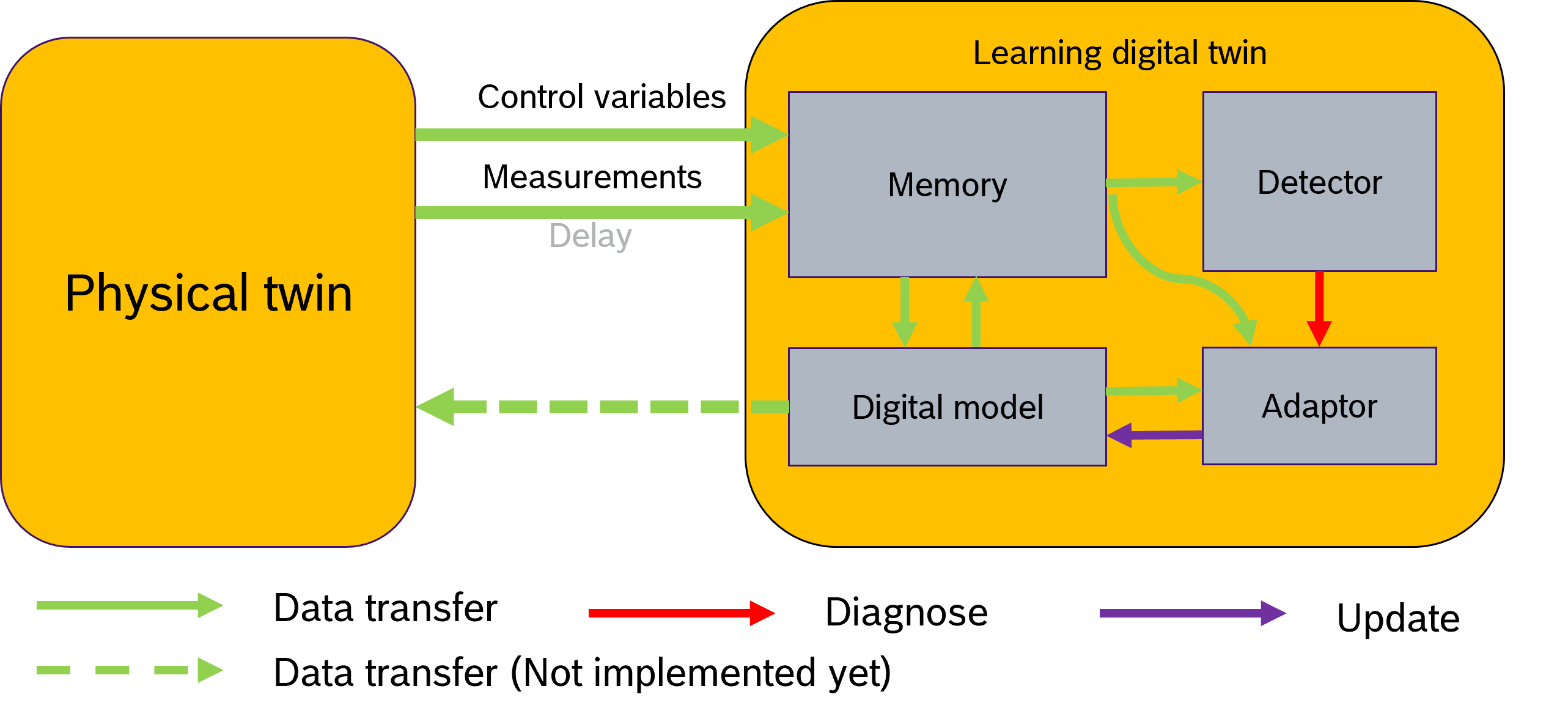}
	\caption{The structure of a LDT and its connection with the physical twin. A LDT consists of four parts: memory, a digital model, a detector, and an adaptor. The backwards influence from LDT to physical twin is not implemented in this paper and is hence denoted by a dashed arrow. } 
	\label{fig:learning digital twin}
\end{figure}

The structure of a LDT is shown in Figure \ref{fig:learning digital twin}. It consists of four parts: memory, a digital model, a detector, and an adaptor. Control variables and measurements are collected from the physical twin and saved in the memory part. The same control variables are used to drive the digital model, which returns the predicted results. The predicted results are also saved in memory. Measurements and predicted results are then compared in the detector, which gives a signal for positive, warning, or negative. A negative signal indicates that the predicted results of the digital model correspond well with the measurements, and hence no adaptation is necessary. A warning signal indicates a slight drift, while an adaptation is, however, unnecessary. The corresponding measurements are labeled as warning. A positive signal triggers the adaptor, which then generates a new digital model based on the current measurement and the warning-labeled measurements.

It should be noted that a data flow from LDT to physical twin is necessary in the definition of DT. However, this backwards influence is not implemented in this paper. This will be added to further research in the future.

%#########################
\subsection{Digital modeling}
\label{sec3_3}
In this work, two different models are identified to describe the dynamical behavior of the physical twin. The first model is a linear model, which can be described as follows:
\begin{equation}
\begin{array}{lcl}
\dot{x} &=& Ax + Bu.\\[6pt]
\hat{y} &=& Cx + Du
\end{array}
\end{equation}

where $A$, $B$, $C$ and $D$ are variable matrices to be identified through collected data. The linear model is identified through \textit{ssest} in MATLAB.

\begin{figure}[h]
	\centering
	\includegraphics[width=1.0\linewidth]{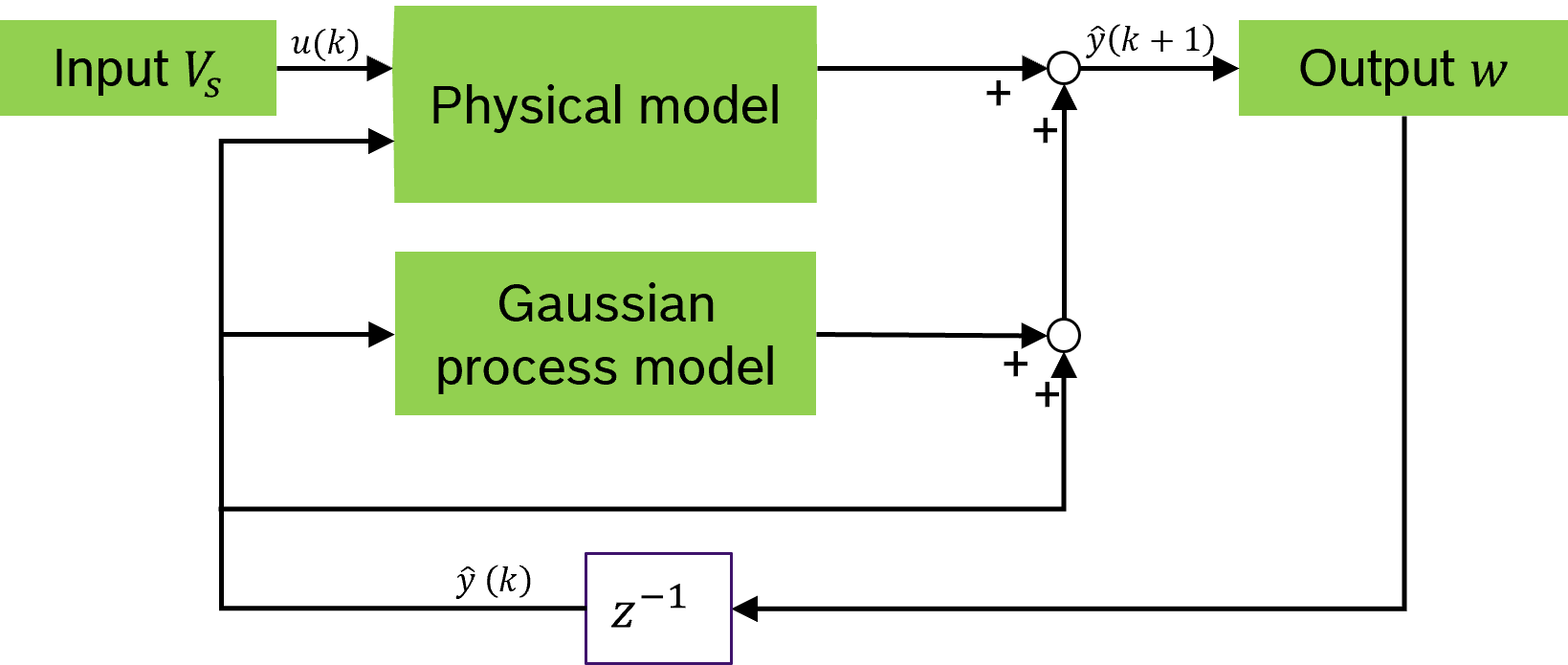}
	\caption{Hybrid model structure.  It consists of two parts, physical model and a data model. In this work, a Gaussian process model is implemented as the data model.} 
	\label{fig:digital modeling}
\end{figure}

The second model implements a hybrid structure, which is shown in Figure \ref{fig:digital modeling}. It consists of two parts: a physical model and a data model. The prediction process of the hybrid model can be described as follows:
\begin{equation}
\begin{array}{cll}
\Delta y &= & f_{p}(u(k), y(k)) + f_{d}(y(k)), \\ [6pt]
\hat{y}(k+1) & = & y(k) + \Delta y, 
\end{array} 
\end{equation}
where $f_{p}$ is the physical model and $f_{d}$ is the data model. In this work, the $y(k)$ is replaced by the prediction $\hat{y}(k)$ because the model works in a closed loop and no measurements are available.
The physical model is an inaccurate and invariant replica of the physical twin and should reflect prior knowledge about it. 
A Gaussian process model was chosen as the data model in this work because of its great performance on approximation and preventing overfitting \cite{b27}. The Gaussian process model is identified  through \textit{fitrgp} in MATLAB.

%##########################
\subsection{Detector}
\label{sec3_4}
An active detector can be described by the following formulation:
\begin{equation}
    \epsilon^i= D(S^i_{p}, S^i_{m}, \theta),
\end{equation}
where $\epsilon^i$ is the output of the detector and $i$ is the index of the current experiment. The value of $\epsilon^i$ is among the set $\epsilon^i \in \left \{1,0,-1 \right \}$, which indicates a positive, warning, or negative signal, respectively.  {\setlength{\baselineskip}{16pt} $S^i_p = \left \{ \hat{y}^i_k \right \}_{k=1}^L$ is the predicted results of the digital model  and $S^i_m = \left \{ y^i_k \right \}_{k=1}^L$ is the measurements in current experiment. $L$ is the length of the experiment. $\theta$ includes all the hyper parameter in the detector. \par} 

Threshold-based detector is the most straightforward detector,  which can be formulated as following:
\begin{equation}
\epsilon^i = \left\{ 
\begin{array}{ll}
1, & \mbox{if } f_m(S^i_p, S^i_m) > \theta_{c}, \\ [6pt]
0, & \mbox{else if } f_m(S^i_p, S^i_m) > \theta_{w}, \\ [6pt]
-1, & \mbox{else } 
\end{array} 
\right.
\end{equation}
where $f_m$ is some metrics to evaluate the difference between the measurements $S^i_m$ and predicted results $S^i_p$. In this paper the  maximum error is chosen as $f_m$.  $\theta_c$ is the threshold value for adaptation and $\theta_w$ is the threshold value for warning, respectively.

Inspired by the drift detection method (DDM) in \cite{b13}, we propose a learning drift detection method (LDDM) for the detection of concept drift in physical twins.  We define an error as when 
\begin{equation}
    abs(y^i_k-\hat{y}^i_k)>\theta_d,
\end{equation}
where $k$ is the time stamp in the $i$-th experiment. $abs()$ is the euclidean  distance between the two items, and $\theta_d$ is the threshold for the evaluation of an error. The error rate $p_i$ refers to the frequency of error in all the experiments since the beginning or last detected drift. 
Similar as in \cite{b13}, $s_i$ is the standard deviation and calculated as 
\begin{equation}
   s_i = \sqrt{\frac{p_i(1-p_i)}{i*L}}, 
\end{equation}
where $i*L$ indicates the summed length of the previous and current experiments. It is assumed that all the experiments are of the same length $L$. Then, the output of LDDM is calculated through following formulation:
\begin{equation}
\epsilon^i = \left\{ 
\begin{array}{ll}
1, & \mbox{if } p_i + s_i \geq p_{min}+3s_{min}, \\ [6pt]
0, & \mbox{else if } p_i + s_i \geq p_{min}+2s_{min}, \\ [6pt]
-1, & \mbox{else  } 
\end{array} 
\right.
\end{equation}
where $s_{min}$ and $p_{min}$ are two registers managed by the LDDM.
As in \cite{b13}, $s_{min}$ and $p_{min}$ are updated whenever $p_i + s_i < p_{min}+s_{min}$. 

%##########################
\subsection{Adaptor}
\label{sec3_5}
According to the output of the detector, different algorithms can be triggered in the adaptor. If $\epsilon^i= -1$, no action is required. If $\epsilon^i= 0$, the current measurements in $S_m^i$ are labeled "warning" and added to the dataset $S_w = S_w \bigcup S_m^i $. If $\epsilon^i= 1$, the adaptation of the digital model is triggered. In this research, the new model $M_{new}$ is obtained by re-identification of the model (linear model) or by retraining the data model (hybrid model) on datasets $S_w$. 

After each adaptation, the warning dataset $S_w$ is reset.
%######################################################################################################
%########################  Simulation results ##########################################################
\section{Simulation Result}
\label{sec4}
In this section, the proposed LDT is implemented to describe the dynamical behavior of a simulated eccentric rotor in the degradation process. It should be noted that although the experiment is conducted on a simulation platform, the workflow for the LDT can be easily applied to a real experiment platform. 

%########################
\subsection{Experiment setting}
\begin{figure}[h]
	\centering
	\includegraphics[width=1.0\linewidth]{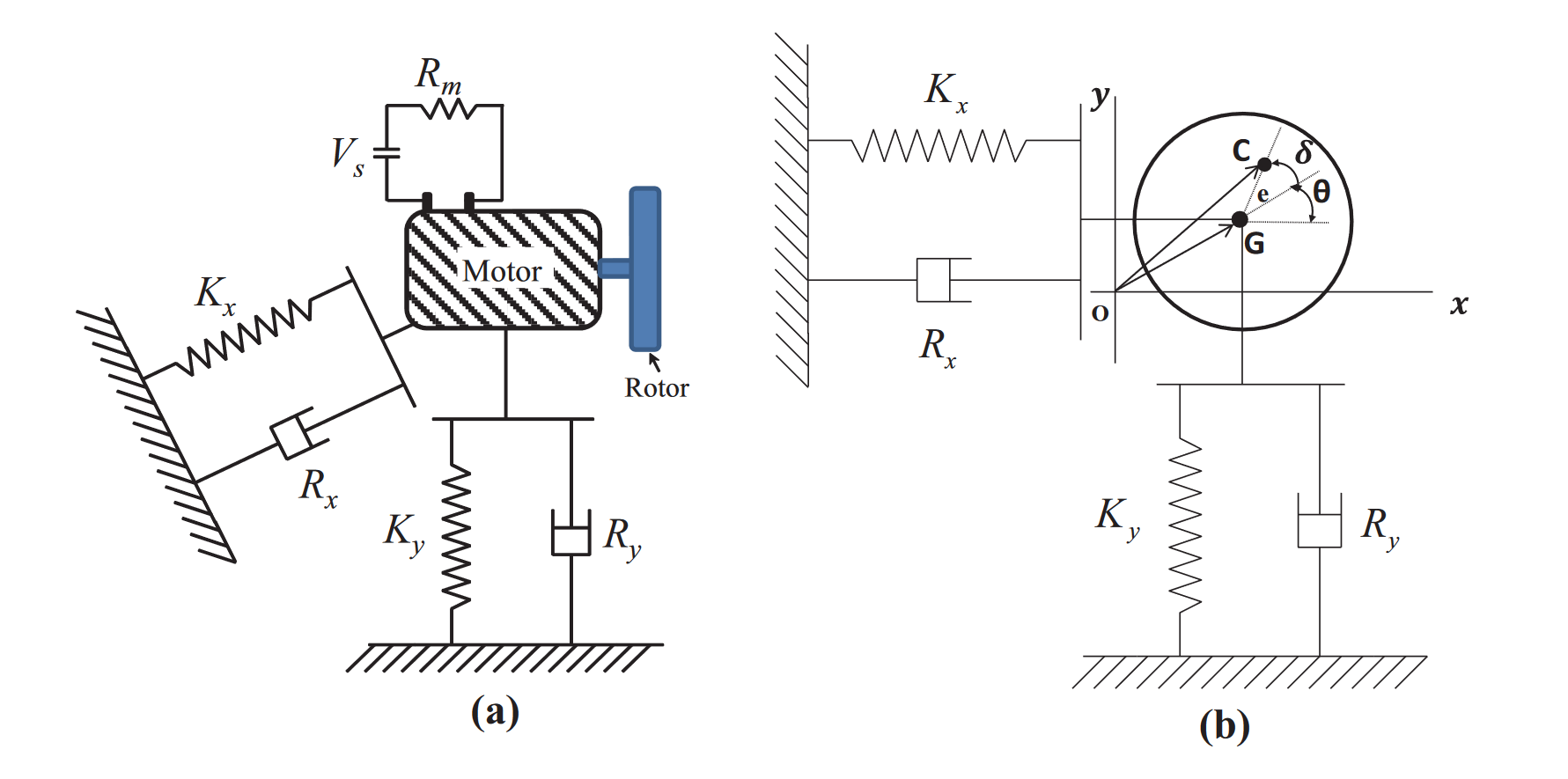}
	\caption{An eccentric rotor driven by a DC motor, which is placed on an anisotropic nonrigid foundation \cite{b22}.} 	
    \label{fig:sommerfeld}
\end{figure}

The simulation platform is adopted in \cite{b22} and shown in Figure \ref{fig:sommerfeld}. A DC motor drives an eccentric rotor. The motor is positioned on an anisotropic, nonrigid foundation. The stiffness and damping of the foundation are denoted as $K_x$ , $K_y$ and $R_x$ , $R_y$, respectively.

\begin{figure}[h]
	\centering
	\includegraphics[width=1.0\linewidth]{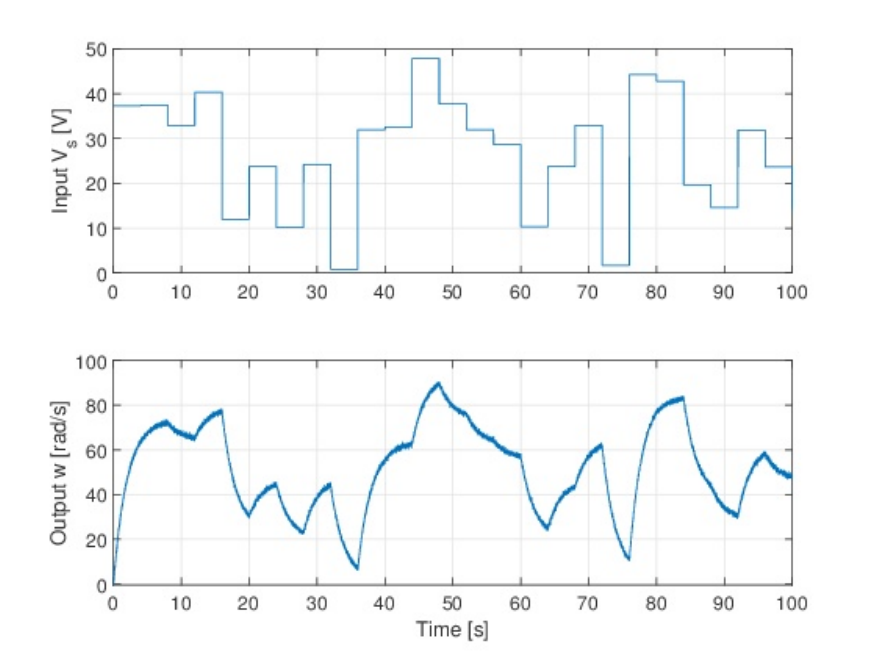}
	\caption{Input and output in an experiment. The supply voltage $V_s$ is chosen as chosen as system input $u$ and the motor velocity $w$ is chosen as output $y$, respectively.} 	
    \label{fig:input_output}
\end{figure}

\begin{figure}[h]
	\centering
	\includegraphics[width=1.0\linewidth]{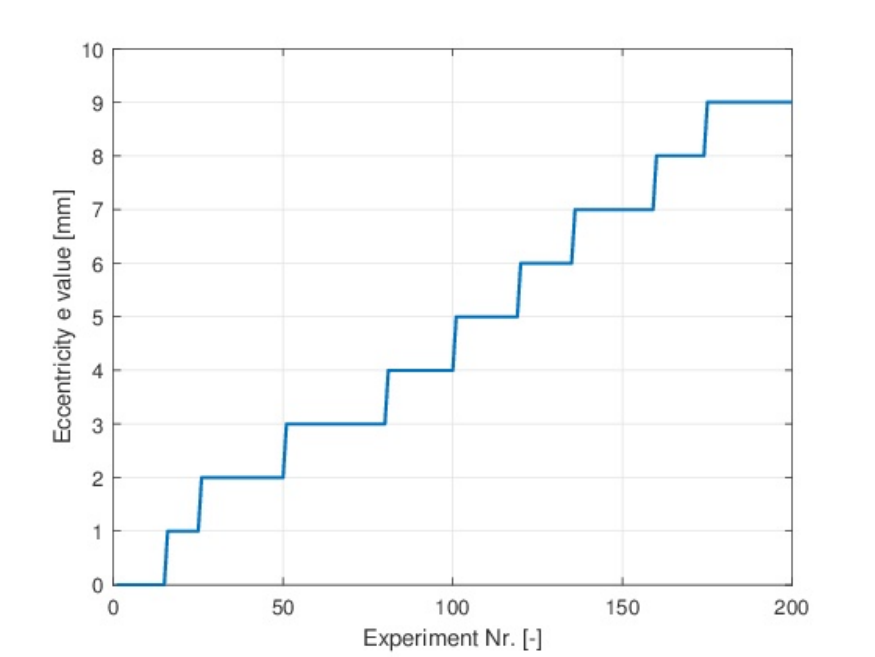}
	\caption{The variation of eccentricity $e$. $e$ increases monotonically as the number of experiments increases. The length of the region in which the $e$ remains stable is randomly generated. It is assumed that the reason for the degradation of the system is unknown and the eccentricity $e$ is not directly measurable.}
    \label{fig:change of e}
\end{figure}

The system is constructed in Simulink. The parameter can be referred to in \cite{b22}. The supply voltage $V_s$ is chosen as the system input $u$, and the motor velocity $w$ is chosen as the output $y$, respectively. The input and output of an experiment are shown in Figure \ref{fig:input_output}. Totally, $N = 200$ experiments are conducted. Each experiment has the same length and sampling rate. In each experiment, $u$ is randomly generated. The degradation of the system is simulated through the variation of eccentricity $e$, which is shown in Figure \ref{fig:change of e}. It can be seen that the $e$ increases monotonically as the number of experiments increases. It is assumed that the reason for the degradation of the system is unknown and the eccentricity $e$ is not directly measurable. The experiments are then provided to LDT in a data stream from Nr. $1$ to Nr. $200$ to simulate the degradation process in reality.  The length of the region in which the $e$ remains stable is randomly generated.

%##########################
\subsection{Result}
\subsubsection{Verification of the learning ability}
\begin{figure}[h]
	\centering
	\includegraphics[width=1.0\linewidth]{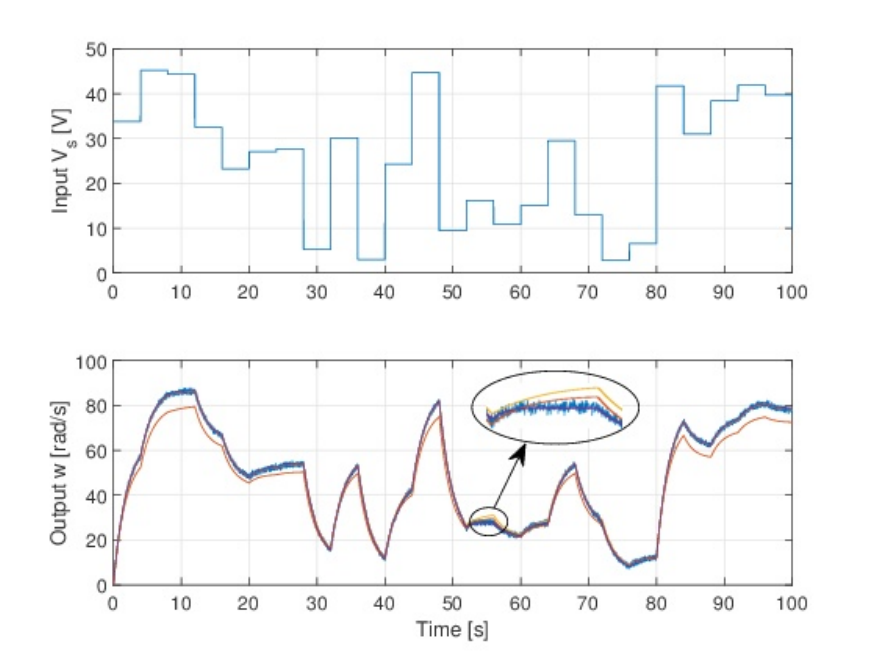}
	\caption{Verification of the learning ability on experiment Nr. $92$. The input $V_s$ is randomly generated. The blue line, red line, yellow line, and magenta line indicate the measurements, the predictions of the original model, the predictions of the linear model, and the predictions of the hybrid model, respectively. It can be seen that both the linear model and the hybrid model are better than the original model. However, only the hybrid model can describe the Sommerfeld effect in certain regions.} 	
    \label{fig:different models}
\end{figure}

As in Figure \ref{fig:different models}, a verification of the learning ability of the LDT is conducted on the experiment Nr. $92$. The original model shows the predicted outputs of an ideal rotor system without eccentricity, which differs from the measurements. It can be seen that in the majority of regions, the trained linear model and the hybrid model provide acceptable accuracy. However, in some regions (as shown in the circle in Figure \ref{fig:different models}), the linear model differs from the measurements. This is the region where the Sommerfeld effect takes place \cite{b22}. The Sommerfeld effect is a nonlinear effect and, hence, cannot be described by a linear model. In comparison, the hybrid model corresponds well with the measurements.

A more detailed comparison of the linear model and the hybrid model can be referred to in Table \ref{Table:model}. We define the mean of maximum errors (MME) as the mean value of the maximum errors in all the experiments. Under the same detector, the hybrid model not only achieves a smaller MME but also smaller adaptation times. We define that a drift is correctly detected when the detector gives a positive signal within the next five experiments after the drift occurs. Recall evaluates the possibility that a drift is correctly detected, and precision evaluates the possibility that a positive detector signal reveals a real drift. Both of the model structures are not well performed on recall. It is because sometimes the drift on the physical twin does not directly lead to  performance degradation. With regard to precision, the hybrid model provides a higher value.

\begin{table}[h]
\caption{Comparison of linear model and hybrid model}
\begin{tabular*}{\hsize}{@{\extracolsep{\fill}}lll@{}}
\toprule
 Metrics & Linear model & Hybrid model\\
\colrule
MME [-] & 8.86 & 6.86       \\
MME of original model [-] & 10.09 & 10.09        \\
Number of adaptation [-] & 44          & 25        \\
Precision [-] & 0.06           & 0.12       \\
Recall [-]   & 0.33         & 0.33        \\ 
\botrule
\end{tabular*}
\label{Table:model}
\end{table}

% As can be seen in Figure \ref{fig:different models}, we can conclude that the learning digital twin algorithm can be applied to both model structures and brings advantages against the model without adaptation. 

%##########################

\subsubsection{Comparison of different detectors}
% A comparison of different detectors on a hybrid model can be referred to in Figure \ref{fig:detector_compare}. As can be seen, all detectors received better performance than the model without adaptation in different experiments. However, all the detectors receive an abrupt increase in error in some experiments. This cannot be avoided since the digital model is optimized for the collected experiments while the error is evaluated for the new experiment. Besides, both threshold-based detector and learning DDM react very quickly to an abrupt increase in error. In comparison, the window-based detector is only activated after a certain time period and hence can be slow under abrupt changes. 

\begin{figure}[h]
	\centering
	\includegraphics[width=1.0\linewidth]{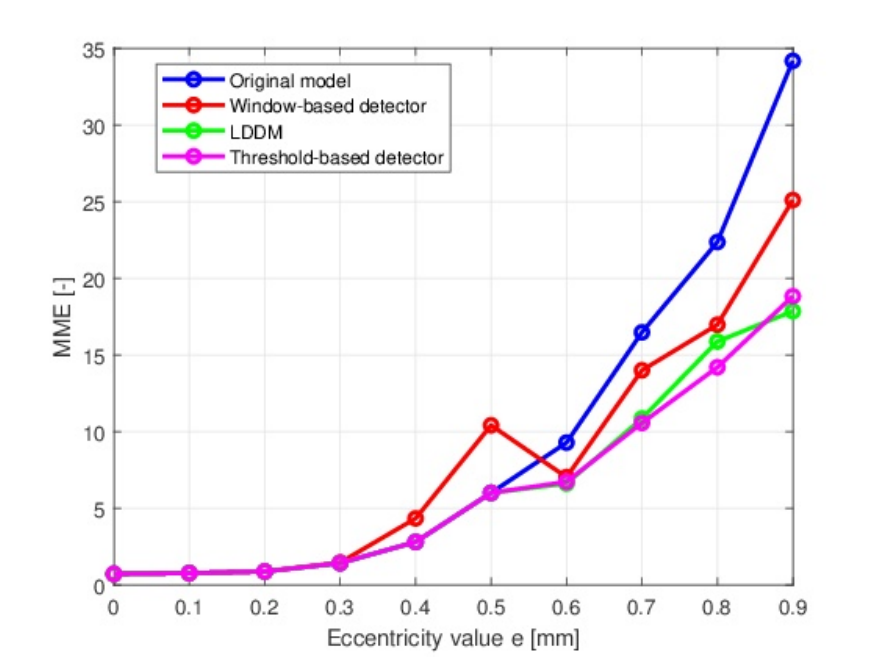}
	\caption{The relationship between the MME and eccentricity value $e$ under different detectors on the hybrid model. The adaptation times of window-based detectors fluctuate with different $e$. In comparison, when $e$ is small, no adaptation is triggered by LDDM and threshold-based detector. More adaptation is triggered when $e$ is large, where the performance degrades fast.} 	
    \label{fig:detector_error}
\end{figure}

Figure \ref{fig:detector_error} shows the relationship between the MME and eccentricity value $e$ under different detectors on the hybrid model. It can be seen that as the eccentricity value $e$ increases, the MME also increases as a result of the Sommerfeld effect. The use of detectors helps decrease the error compared to the original model. In comparison with the window-based detector, the LDDM and the threshold-based detector can achieve a smaller MME.

\begin{figure}[h]
	\centering
	\includegraphics[width=1.0\linewidth]{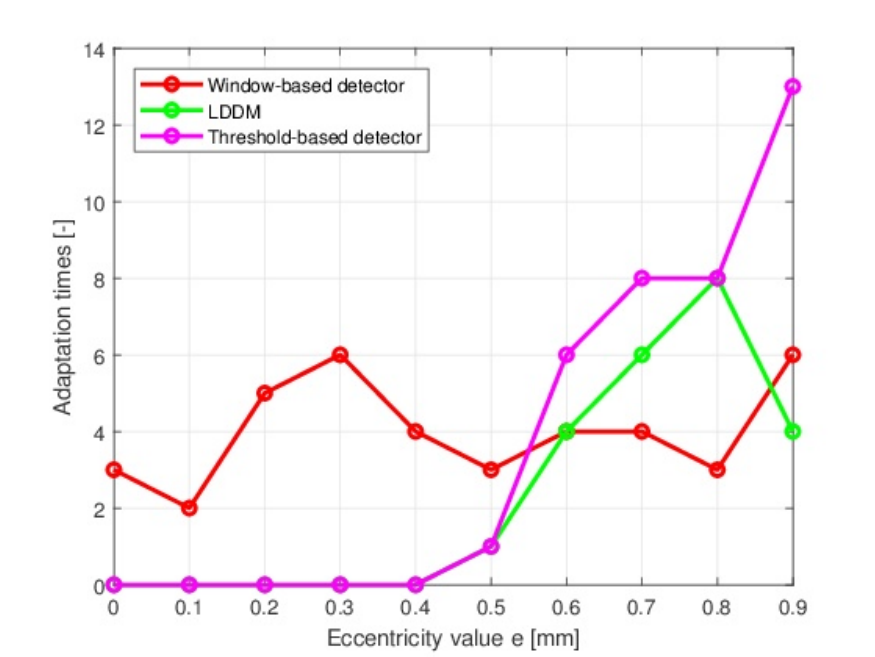}
	\caption{ The relationship between the adaptation times and eccentricity value $e$ under different detectors on the hybrid model. All detectors received a smaller error than the model without adaptation in different experiments. A threshold-based detector and LDDM are more sensitive to an abrupt increase in error than a window-based detector.} 	
    \label{fig:detector_adaptation}
\end{figure}

Figure \ref{fig:detector_adaptation} shows the relationship between the adaptation times and eccentricity value $e$ under different detectors on the hybrid model. Because the length of the region in which the $e$ remains stable is randomly generated, the adaptation times of window-based detectors fluctuate with different $e$. In comparison, the threshold-based detector and LDDM only give a positive signal when performance degradation is detected. Hence, when $e$ is small, no positive signal is given. More adaptation is triggered when $e$ is large, where the performance degrades fast. This explains why LDDM and threshold-based detectors achieves better performance on MME. 

%##########################
\subsubsection{Influence of threshold value on detectors}
\begin{figure}[h]
	\centering
	\includegraphics[width=1.0\linewidth]{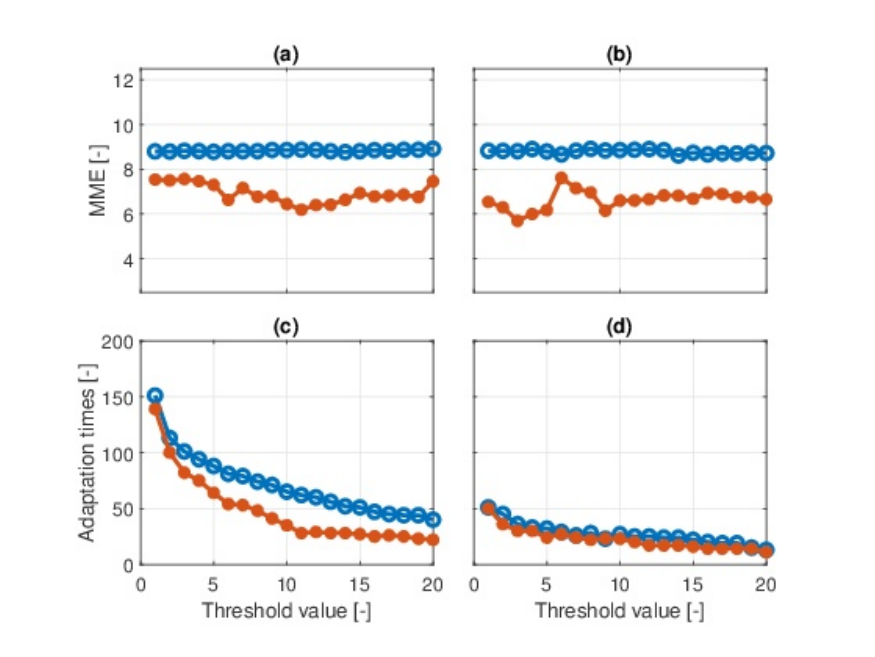}
	\caption{The influence of threshold value on detectors. The red line and blue line indicate the results of the hybrid model and the linear model, respectively. (a) MME with different threshold values in a threshold-based detector; (b) MME with different threshold values in LDDM; (c)Adaptation times with different threshold values in a threshold-based detector; (d)Adaptation times with different threshold values in LDDM.} 	
    \label{fig:threshold_influence}
\end{figure}

% \begin{figure*}[t]\vspace*{0.6pt}
% %\centerline{\includegraphics{fx1}\hspace*{5mm}\includegraphics{fx1}}
% \centerline{\includegraphics[scale=0.8]{threshold_influence.eps}}
% \caption{(a) Summed error with different threshold values in a threshold-based detector; (b) Summed error with different threshold values in learning DDM; (c)Adaptation times with different threshold values in a threshold-based detector; (d)Adaptation times with different threshold values in learning DDM.}
% \label{fig:threshold_influence}
% \end{figure*}
Figure \ref{fig:threshold_influence} shows the influence of the value of threshold on adaptation times and MME in threshold-based detector and LDDM. In this experiment, the $\theta_w$ is set to be $1/2*\theta_c$ in order to reduce variables. It is first to be noted that the hybrid model achieves better performance than the linear model under different threshold values. In both detectors, a smaller threshold leads to greater adaptation times for the linear model and the hybrid model. The MME of a linear model does not show a correspondence with different threshold values. This is because the linear model has a greater structural error because of the Sommerfeld effect. Hence, although the adaptation time varies, the summed error stays stable. In the following, we focus on analyzing the influence of threshold values on the hybrid models under different detectors.

In a threshold-based detector, it can be seen that the MME achieves its minimal value in the middle of the range of different threshold values. When the threshold value is too large, the digital model becomes insensitive to the change in the physical twin. In contrast, the MME also increases when the threshold value is small,  although the adaptation time increases greatly. This may be because the model is poorly trained because of the high frequency of adaptation.

In LDDM, a fluctuation of MME under different threshold values can be observed. In comparison with the threshold-based detector, the adaptation times under LDDM are much smaller at the same threshold values. This is meaningful because, generally, in reality it is difficult to fine-tune a suitable threshold value. In order to save effort, the adaptation times should be kept in an acceptable region. This indicates that the LDDM is a more robust detector considering the adaptation times.

%###############################################################################################
\section{Conclusion and outlook}
\label{sec5}
This research presents a LDT for the simulation of the temporal behavior of a physical object. The learning ability of the LDT increases the accuracy of the predicted result when the physical object suffers performance degradation. The structure of a LDT consists of four parts: memory, a digital model, a detector, and an adaptor. The proposed LDT is verified on the simulated degradation process of an eccentric rotor system. Both the linear model and the hybrid model are implemented in the LDT and receive greater accuracy in comparison with the original model without adaptation. The learning ability is realized through re-identification of the model (linear model) or by retraining the data model (hybrid model). Different detectors are constructed and compared in the LDT. It can be concluded that the use of active detectors increased the learning performance compared to passive detectors. In addition, LDDM has a greater robustness against a threshold-based detector considering the resulted adaptation times. In the future, research will be carried out on how to find the optimal hyperparameters for the detectors.

%##########################  End   
%#####################################################################################################

% \bibitem[Deal and Grove(2009) ]{Deal}Deal, B., Grove, A., 1965. General Relationship for the Thermal Oxidation of Silicon. Journal of Applied Physics 36.2, 37--70.

% }Fachinger, J., den Exter, M., Grambow, B., Holgerson, S., Landesmann, C., Titov, M., Podruhzina, T., 2004. ``Behavior of spent HTR fuel elements in aquatic phases of repository host rock formations,'' 2nd International Topical Meeting on High Temperature Reactor Technology. Beijing, China, paper \#B08. 

\end{document}